\documentstyle[aaspp4,12pt, psfig]{article}

\newcommand{\be}{\begin{equation}}
\newcommand{\ee}{\end{equation}}
\newcommand{\nn}{\mbox{} \nonumber \\ \mbox{} }
\newcommand{\ba}{\begin{eqnarray}}
\newcommand{\ea}{\end{eqnarray}}
\newcommand{\om}{\omega}
\newcommand{\Alfven}{Alfv\'{e}n }
\def\lo{\mathrel{\raise.3ex\hbox{$<$}\mkern-14mu\lower0.6ex\hbox{$\sim$}}}
\def\go{\mathrel{\raise.3ex\hbox{$>$}\mkern-14mu\lower0.6ex\hbox{$\sim$}}}

\begin{document}

\title{Dynamics of relativistic reconnection}

\author{Maxim Lyutikov  $^{1,2,3}$}
\author{Dmitri Uzdensky $^{4}$}

\affil{$^1$ Physics Department, McGill University, 3600 rue University
Montreal, QC,  Canada H3A 2T8, \\
$^2$ Massachusetts Institute of Technology,
77 Massachusetts Avenue, Cambridge, MA 02139, \\
$^3$ CITA National Fellow}
\affil{ $^4$  Kavli Institute for Theoretical Physics, 
University of California, Santa Barbara, CA 93106}

\date{Received   / Accepted  }

\begin{abstract}
The dynamics of the 
 steady-state Sweet--Parker-type reconnection is analyzed
in relativistic regime when  energy
density 
 in the inflowing region 
 is dominated by  magnetic field.
The structure of  reconnection layer (its thickness, inflow
and outflow velocities) 
depends on the ratio of  two large
dimensionless parameters of the problem - magnetization parameter
 $\sigma \gg 1$
(the ratio of the magnetic to particle energy-densities in the
inflowing region)
and the Lundquist number $S$. The  inflow velocity
may be relativistic (for $ S< \sigma$) or non-relativistic
 (for $ S > \sigma$), while the outflowing plasma
is moving  always relativisticly.
For extremely  magnetized plasmas with 
$\sigma \geq S^2$,  the inflow four-velocity becomes of the order
of the \Alfven  four-velocity.
\end{abstract}

\section{Introduction}

Magnetic reconnection is widely recognized as a very important phenomenon
in many laboratory and astrophysical plasmas (Biskamp 2000,
Priest \& Forbes 2000). It has been
studied very extensively over the last 40 years, and a very 
significant progress has been made in understanding this
process. However, historically, reconnection was of interest mostly
to space physicists studying the Solar corona and the Earth's 
magnetosphere and to researchers in  magnetic confinement fusion. In  all
these environments,  plasma flows are non-relativistic
and the  \Alfven velocity is usually much less than
the speed of light (equivalently, magnetic energy density is much
smaller than the particle rest mass  energy density).
Therefore it is not surprising that most of the progress on the subject 
has been made in  non-relativistic regime.

Over the last 
decade however, it has been recognized that magnetic reconnection 
processes are also of great importance in high energy astrophysics,
 where  dynamic behavior
is often dominated by  super-strong magnetic fields,
with  energy density $B^2/(8 \pi)$ larger than 
the rest energy of the matter
$\rho c^2 +\epsilon$. 
The best studied (but yet not completely understood) case is  
magnetized winds from pulsars. 
 Models of pulsar 
magnetosphere (Goldreich \& Julian 1969,
 Arons \& Scharlemann 1979, Ruderman \& Satherland 1975) 
 predict that  near the light cylinder  most of the spin-down luminosity of  
a pulsar 
 should be in a form of Poynting flux.
Other possible examples  of relativistic strongly magnetized
media
include  jets emanating from magnetized accretion disks around
Galactic black holes and neutron stars as well as 
Active Galactic
Nuclei (e.g., Beskin 1997, Lovelace et al. 2002), 
magnetosphere of magnetars (Thompson \& Duncan 1996, Thompson et al. 2002)
  and  Gamma Ray Bursters (Lyutikov \& Blandford 2002). 

Dissipation of such super-strong  magnetic
fields
may play an important role both for the global dynamics of the 
system and as a way to produce  high energy emission. 
Magnetic reconnection
 has been proposed as the mechanism for 
acceleration of  pulsar winds (Coroniti
 1994, Kirk \& Lyubarsky 2001) and GRB outflows  (Spruit et al. 2001),
and as a dissipation mechanism in 
 AGN jets (Romanova \& Lovelace 1992), Soft Gamma Ray Repeaters 
(Thompson \& Duncan 1996), 
GRBs (Lyutikov \& Blandford 2002, Spruit et al. 2001).
In case of pulsar winds there are strong arguments that
effective dissipation of magnetic field is, in fact, {\it needed}
to account for the global  dynamics of the Crab  nebula
(Kennel \& Coroniti 1984; see also Michel 1994, Coroniti 1990,
 Melatos \& Melrose
1996,
 Lyubarsky \& Kirk 2001). 

 This provides the motivation for
studying magnetic reconnection in  strongly relativistic plasmas
(to be defined below). Despite of the growing interest in relativistic 
magnetic reconnection, very little theoretical (let alone experimental!)
work has been done on the subject so far. We are aware of only one
 analytical discussion of  relativistic reconnection (Blackman \& Field 1994)
and two recent numerical works on particle dynamics in 
relativistic
reconnection layers (Zenitani \& Hoshino 2001, Larrabee et al. 2002).

In any theoretical analysis of magnetic reconnection,
one first makes a number of approximations,
e.g. incompressibility, two-dimensionality,
the absence (or presence) of the axial magnetic field, etc.
Then one formulates the set of MHD equations in
a dimensionless form where the relative importance
of various physical processes is represented
by certain dimensionless parameters.  After that one then
tries to build a qualitative description of the reconnecting
system by a small number of (also dimensionless)
characteristic ratios. For example, in the simplest
Sweet-Parker model of reconnection (e.g., Priest \& Forbes 2000)
 one assumes
incompressibility, uniform and constant resistivity $\eta$,
and so on, and one finds that  the principal dimensionless
parameter governing the system's behavior is
the Lundquist number $S\equiv V_A L/\eta\gg 1$, where
$V_A$ is the upstream \Alfven velocity and $L$ is the size of the system.
The other dimensionless plasma parameter, $\beta$ - the ratio of
thermal pressure to magnetic field energy density, 
turns out not to be  as important, at least in the first approximation.
Correspondingly $S^{-1}$ plays a role of the small parameter
on which all further asymptotic expansions and boundary layer
analysis are based. One then seeks to find out how the dimensionless
characteristics of the system, such as the reconnection layer's
aspect ratio $\delta/L$ and the ratio of the incoming velocity
 to the \Alfven velocity 
scale with $S$ as $S\rightarrow \infty$.

The generalization of reconnection to the relativistic case  requires
 an introduction of one more (in addition to $S$) principal
dimensionless parameter that should describe how far in 
relativistic regime we are. This parameter is $\sigma$, the
ratio of the magnetic field energy density to the plasma's
rest mass energy density in the inflow region.
We are now in position to define what we mean by relativistic reconnection:
 we consider a case when   magnetic energy density
in the flowing plasma dominates over particle energy density,
$\sigma \gg 1$. During reconnection   magnetic energy is
dissipated and transformed into  plasma thermal energy and later
into bulk motion. Since $\sigma \gg 1$,  energy per baryon becomes much
larger than $m_p c^2$  and so we get relativisticly hot
plasma in the reconnection layer. Expansion of  plasma along the reconnection
 layer will produce relativistic bulk motions downstream.
In  relativistic regime, $\sigma \gg 1$,  the
\Alfven velocity becomes relativistic,
  $V_A = \sqrt{ \sigma/(1+\sigma)} c \simeq c$, so that 
 under certain conditions (to be determined later)
the upstream flow is expected to be relativistic as well.

Thus, in case of relativistic reconnection 
there are two very large dimensionless parameters, $S$
and $\sigma$. As we shall see below, one gets two different
regimes depending on the ratio of these parameters: in one
regime the incoming (the upstream) flow is ultra-relativistic,
while in the other it is non-relativistic.

\begin{figure}[h]
\psfig{file=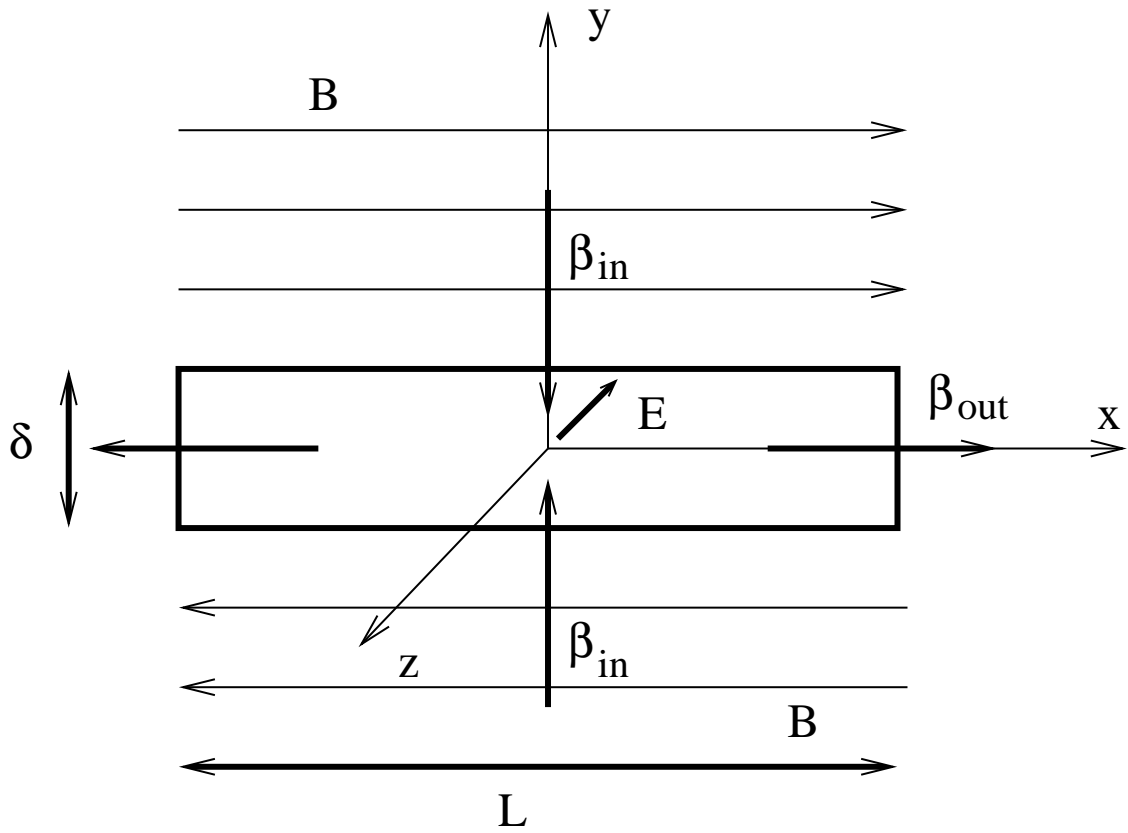,width=17cm}
\caption{Geometry of the model. The reconnection layer
has a width $L$ in $x$ direction
 and thickness $\delta$  in $y$ direction (and infinite depth
in $z$). Outside  there is strongly magnetized plasma
flowing into  the  reconnection layer with velocity $\beta_{\rm in}$
along the $y$ direction.  Inside the magnetic energy is converted into
internal energy and the flow is accelerated along the
$x$ axis to outflowing velocity $\beta_{\rm out}$.
}
\label{1}
\end{figure}

In this paper we present a relativistic generalization of the 
simplest model of magnetic reconnection --- the Sweet--Parker
model --- to strongly relativistic plasmas. This is a simple 
two-dimensional resistive MHD model, presented in Figure \ref{1}.
We assume that the reconnection layer has a rectangular
shape with  a width $L$ and thickness
$\delta \ll L$. 
The width $L$ of the reconnection layer is determined
by the global system size and thus, for the purposes of studying
magnetic reconnection, is a fixed prescribed quantity.
Also prescribed are the magnetic field strength
and the baryon density and pressure in the ideal-MHD inflow region
above and below the reconnection layer. In contrast, the thickness
$\delta$ of the reconnection region, as well as some other parameters
such as the plasma inflow and outflow
 velocity, are not prescribed and need to be
calculated as a part of the analysis.

We basically follow the steps of the Sweet--Parker
analysis while taking into
 account relativistic effects, such as relativistic
contraction and  inertia of magnetic field.
As the plasma enters the reconnection layer
it slows down coming to a halt at  stagnation point.
At the same time the magnetic energy is dissipated and converted into
internal energy of  pair-rich plasma.
In the out-flowing region the plasma is accelerated by the 
pressure gradient in $x$ direction reaching some terminal relativistic velocity
$\gamma_{\rm out}$.
We assume that  energy losses are not important
and the  total energy of a fluid element (or at least a
large fraction)  stays
within this fluid element, providing the corresponding amount
of pressure support.  Though generally 
radiative losses may  be  important, we expect that the 
reconnecting plasma will be optically thick to Thomson scattering
after it's temperature  becomes weakly relativistic, $T \geq 20$
keV. Above this temperature an efficient
pair production process will start (e.g., Goodman 1986) inside the
reconnection current layer trapping the radiation.
 The role of pair production
in increasing the optical depth is an interesting
question in itself and deserves further study 
(Thompson 1994). It lies, however, outside the
scope of our paper; here we simply assume that
the plasma is optically thick inside the reconnection
layer  and so the released
magnetic field energy cannot leave the system and
is therefore available for accelerating plasma downstream.

\section{Relativistic reconnection formulation}

The basic equations include  the relativistic Ohm's law
and the 
relativistic dynamics, Maxwell's and mass conservation equations
 (Lichnerowicz 1967):
\ba &&
T^{ij}_{,i} =0,
\label{x1}
 \\ &&
F^{\ast\, ij}_{,i} =0,
\label{x2}
\\ &&
(\rho u^i)_{,i}=0
\label{x3}
\ea
where
\be
T^{ij} = (w +b^2+\epsilon^2) u^i u^j +(p+ {b^2+ \epsilon^2 \over 2}) 
g^{ij} -b^i\,b^j - \epsilon^i \epsilon^j
\label{T}
\ee
is the stress-energy tensor,
$w $ is  the
 plasma proper enthalpy,
 $\rho$ is proper plasma density and $p$ is pressure,
$b^2 = b_i b^i$ and $\epsilon^2 =  \epsilon^i  \epsilon_i$  
are
 the plasma proper magnetic and electric energy density times $4 \pi$,
$p$ is pressure, $u^i=(\gamma, \gamma {\bf \beta})$
are the plasma four-velocity,
Lorentz-factor and three-velocity, $g^{ij}$ is the metric tensor,
$b_i = {1\over 2} \eta_{ijkl} u^j F^{kl}$ are the four-vector of magnetic
field, Levy-Chevita tensor and electro-magnetic field tensor and
$\epsilon_i = u^j F_{ji}$ is the four-vector of the electric field.

The choice of the stress-energy tensor deserves some discussion.
The stress-energy tensor (\ref{T}) is a full
electro-magnetic field plus matter tensor; this is  {\it not}
the relativistic magneto-hydrodynamic (RMHD) stress-energy tensor.
 The reason is that
in the frame work of RMHD it
 is  assumed  that one of the electro-magnetic invariants is not
equal to 0 and electro-magnetic stress energy tensor can be
diagonalized.  Equivalently,  this implies that there
is a  reference frame where  electric  field is  equal to
0. In case of resistive  RMHD  such  frame may not exist, 
since,
generally,
there are  resistive electric fields  in the plasma rest-frame
either along the null magnetic line (this violates then the $B^2-E^2 >0$ 
condition of the ideal MHD),
 or field aligned electric fields (this  violates
${\bf E}\cdot {\bf B}=0$ condition).
In the full relativistic approach the  resistive electric field 
contributes to plasma energy density, energy fluxes and stresses.
Nevertheless, we can still define the plasma rest frame by requiring that
in that frame the electric fields $\epsilon$ are only of  resistive nature.
We then obtain the stress-energy tensor (\ref{T}).

In what follows we will be using both the rest frame quantities
(${\bf b} ,\,  {\bf \epsilon},\, p, \, w,\,
\rho_e^{\ast},\, \rho^{\ast}$ for renormalized
magnetic and electric fields, pressure enthalpy and charge and mass densities
 as well as laboratory quantities ${\bf B} ,\,  {\bf E} ,\,
\rho_{e}  ,\, \rho $ (pressure and enthalpy are defined only 
in the rest frame).

\subsection{Relativistic Ohm's law}

Relativistic 
Ohm's law is
\be
j^j = (ju) u^j + {1 \over \eta} F^{i k} u_k
\label{Oh}
\ee
where $\eta$ is the plasma resistivity (e.g.  Lichnerowicz 1967).
In 3-D notations (Greek indexes =1,2,3)
this gives
\ba &&
j^0 \equiv \rho_{e}= { ({\bf j} \cdot {\bf \beta}) \over \beta^2}   -  { 1
 \over \eta \gamma}
( {\bf  E}  \cdot {\bf \beta})
\nn &&
{\bf j} = \left( j^0 -  ({\bf j} \cdot {\bf \beta}) \right)
  \gamma^2  {\bf \beta}+  {\gamma
 \over \eta}
\left( {\bf  E} + ( {\bf \beta} \times  {\bf B})  \right)
\label{j}
\ea
which can be written
\be
\left(  \delta^{\alpha \beta} - {  \beta^\alpha  \beta^\beta \over \beta^2}
  \right)
\left( j_ \beta -   {\gamma
 \over \eta}  E_ \beta \right) =
  {\gamma
 \over \eta}  \left.( {\bf \beta} \times  {\bf B})\right|^\alpha
\label{Ohm}
\ee
The form (\ref{Ohm}) of the relativistic Ohm's law 
 shows that for $({\bf j} \cdot {\bf \beta}) =0$
and $( {\bf  E}  \cdot {\bf \beta}) =0$ the relativistic effects
change the conductivity
\be 
 \eta \rightarrow { \eta \over \gamma}
\ee
This can be understood if one notices that in the plasma rest frame
the electric field $\epsilon=E/\gamma$, and, since for 
$({\bf j} \cdot {\bf \beta}) =0$
the current in the rest frame is ${\bf j}$
\be
{\bf j} ={ \epsilon \over \eta}
\ee

\subsection{ Main equations}

For stationary flow the 
energy  and momentum flux conservation can be rewritten
\ba &&
\nabla_\alpha  \left(  \gamma^2 (w+ b^2) \beta   \right) =0
\nn &&
\nabla_\beta  \left(  \gamma^2 w \beta ^ \alpha \beta ^\beta - p \delta^{  \alpha \beta} \right) =
F^{ \alpha \beta} j_ \beta =
 \rho_e E^\alpha + \left.(  {\bf j} \times  {\bf B})\right|^\alpha
\ea
The Maxwell's equations then become
\ba &&
{\bf j}  = {1\over 4 \pi} \nabla  \times  {\bf B} 
\nn &&
 (\nabla  \times  {\bf E})=0
\nn &&
{\rm div} {\bf E} = 4 \pi \rho_e
\label{M}
\ea 
The equation of continuity is
\be
{\rm div} \gamma  {\bf \beta } \rho^\ast =0
\ee
The above equations plus 
 the equation of state form a system of 15 equations for 15 variables
$ {\bf \beta },  {\bf E} ,  {\bf B},{\bf j}, \rho_e,  \rho, p$.

Following the Sweet--Parker model we neglect  possible field-aligned
electric fields and currents. Then the only non-vanishing components of
electric field and  current are $E_z$ and $j_z$. Since the velocity
lies in the $x-y$ plane from eq. (\ref{j})  it follows that the
charge density $\rho_e\equiv  j^0 =0$. 
This is an important simplification
since generally 
 for relativistic plasma the charge density can not be neglected.

\subsection{ Magnetization parameter}

We assume that far in the incoming region  plasma is cold and strongly 
magnetically dominated. In the  incoming 
region, well outside the reconnection layer, the restive
electric field is vanishing and
 can be neglected.  We introduce a frame-invariant 
 ratio of the rest energy
density of magnetic field and particles 
\be 
\sigma = {b_{\rm in}^2 \over \rho^{\ast}} \gg 1
\ee
This definition of $\sigma$ is equivalent to the 
ratio of Poynting to particle fluxes in the incoming region
(c.f., Kennel \& Coroniti 1984).
 For example, for  pulsar winds initially (near the light cylinder)
$\sigma \sim 10^3 - 10^6$ (e.g., {Arons} \& {Scharlemann} 1979).

We have defined the $\sigma$-parameter in a frame-invariant way in
 terms of the rest-frame quantities.
Alternative definition involves the laboratory frame fields and 
densities
\be 
\sigma_l={B_{\rm in}^2 \over  \rho} =  \gamma_{\rm in} 
  \sigma 
\ee
It is straightforward to express the results in terms of
$\sigma_l$ instead of $\sigma$.

\section{Flow along the velocity separatrix}

In non-relativistic reconnection an important model problem is the flow
of plasma along the velocity separatrix $x=0$ (e.g., Priest \& Forbes 2000).
In this case a simple equation relating  magnetic field strength and
inflow velocity can be obtained using only Ohm's law (\ref{Ohm}). 
Solving this equation for a given velocity (or vise versa for a 
given magnetic field)  profile would allow us to find an example 
of velocity and magnetic distributions. 

The relativistic Ohm's law~(\ref{Ohm}) along the velocity 
separatrix can be written as
\be
j(y) \equiv j_z(x=0,y) ={{\gamma(y)}\over\eta} \, 
(E_z - \beta_y B_x) \, .
\ee

For a more compact notation we introduce $B(y)\equiv B_x(x=0,y)>0$ 
and also $\beta(y)\equiv-\beta_y(x=0,y)>0$ (here we changed the sign 
so that~$\beta$ is positive for convenience). 
Then, the above equation for Ohm's law can be rewritten as  
\be
j(y) \equiv j_z(x=0,y) ={{\gamma(y)}\over\eta} \, 
[E_z + \beta(y) B(y)] \, .
\ee

In a steady case Maxwell's equation gives  $\nabla\times E = 0$, and hence,
using $\partial /\partial z = 0$,
\be
E_z(x,y) = {\rm const} \equiv E \, .
\ee
In the ideal region above the layer, Ohm's law gives
\be
E = - \beta_{\rm in} B_{\rm in} \, ,
\ee
where $B_{\rm in} \equiv B(y\gg\delta)$, and 
$\beta_{\rm in} \equiv \beta(y\gg\delta)$.

Using the first of the  Maxwell equations (\ref{M})
in laboratory frame, we find
\be
\eta {{\partial B}\over{\partial y}} = - \gamma (E + \beta B)\, ,
\ee
or, using our expression for $E$,
\be
{\eta\over\gamma} \partial_y \hat{B} = \beta_{\rm in} - \beta\hat{B}\, ,
\label{B}
\ee
where $\hat{B}\equiv B/B_{\rm in}$. This equation should be 
supplemented by the boundary conditions $\hat{B}(y=0)=0$, 
$\beta(y=0)=0$ and $\hat{B}(y \rightarrow \pm \infty)=\pm 1$,
$\beta(y \rightarrow \pm \infty)= \pm \beta_{\rm in}$.
Equation (\ref{B}) relates the two functions ---  inflow velocity and 
magnetic field along the separatrix. For example, for a given $\beta(y)$
the two boundary conditions for the first-order ODE (\ref{B}) 
determine the solution $\hat{B}(y)$ and  put a constraint 
on the parameters $\eta$, $\beta_{\rm in}$, and~$\delta$.

For example, eq. (\ref{B}) can be resolved for $\hat{B}(y)$ for a given
$\beta(y)$:
\be
\hat{B}(y) \equiv {B(y) \over B_{\rm in}}  = {\beta_{\rm in}  \over \eta} 
 e^{-  { \int^y \beta(y') \gamma(y') dy' \over \eta} }
\int^{y}   e^{ { \int^{y'} \beta(y^{\prime \prime}) 
\gamma(y^{\prime \prime}) dy^{\prime \prime} \over \eta} }
\gamma(y') dy' \, .
\ee
The boundary condition $\hat{B}(y=0)=0$ is automatically satisfied, while
the condition $\hat{B}(y=\delta) =1$ serves as an eigenvalue
problem for $\delta(\eta,\beta_{\rm in})$.

As an example, consider a case where the inflow four-velocity,
$u \equiv \gamma \beta$, is a linear function of distance along
the $y$~axis:
\be
u=\left\{
\begin{array}{ll}
u_{\rm in} { y \over \delta} & \mbox{ if $|y|< \delta$} \\
u_{\rm in}  & \mbox{ if $|y| > \delta$}
\end{array} \right.
\label{u}
\ee
By definition, $\delta$ is the scale on which the inflow velocity changes
from initial $u= u_{\rm in}$ to zero.

Introducing dimensionless parameter $Y \equiv \delta/\eta$ ($Y=c\delta/\eta$
in dimensional  units) and rescaling coordinate~$y$ by~$\delta$, 
$\tilde{y} = y/\delta$, the magnetic field is then determined by
\be
{B\left( \tilde{y} \leq 1 \right) \over B_{\rm in} } 
= { u_{\rm in} Y  \over \sqrt{1+u_{\rm in}^2 } }
 e^{ - u_{\rm in} \tilde{y}^2 Y /2 } 
\int_0^{ \tilde{y}} e^{  u_{\rm in} \tilde{y}^2 Y /2  }
\sqrt{ 1 +  {  \tilde{y} } ^2  u_{\rm in}^2 } d\tilde{y}
\label{wK}
\ee
Parameter $Y$ here is an implicit function of $u_{\rm in}$ given by
the condition $B( 1)= B_{\rm in}$. Numerical solution $Y(u_{\rm in})$ 
is  plotted in Figure \ref{2}.
The corresponding  magnetic field profiles for different
values of $u_{\rm in}$ are plotted in Figure \ref{3}.

Several important conclusions can be drawn from this exercise: 
(i) both the four velocity and magnetic field have {\it the same} 
typical scale $\sim\delta$;
(ii) both in the non-relativistic ($u_{\rm in}\ll 1$) and strongly 
relativistic ($u_{\rm in}\gg 1$) regimes the ratio $Y=c\delta/\eta$ 
asymptotically becomes inversely proportional to the four-velocity 
of the incoming flow, $Y \sim 1/u_{\rm in}$, in agreement with the 
Sweet--Parker theory (in the non-relativistic case). Asymptotic scalings
of $Y(u_{\rm in})$ for $u_{\rm in}\ll 1$ and $u_{\rm in}\gg 1$ are 
analyzed in Appendix \ref{Yassym}.
(iii)
in the case of strongly relativistic inflow, there
is a thin sub-layer near the neutral point, $y=0$,
with the thickness $\delta_{\rm nr} \sim \delta/u_{\rm in}$, 
where the flow becomes non-relativistic.%
\footnote
{Because of the very large proportionality coefficient between~$Y$
and $u_{\rm in}$ in the strongly-relativistic case, this sub-layer 
develops only for very large (of order $10^2$ and greater) values 
of~$u_{\rm in}$, as can be seen from Figure~\ref{3}.}
Outside of this sub-layer, the function $\hat{B}(\tilde{y})$ 
approaches a universal shape in the limit $u_{\rm in}\rightarrow\infty$. 
Inside the sub-layer the magnetic field becomes linear,
\be
B \sim { y  \over \eta} B_{\rm in}, \mbox{ for $ y \ll \delta/u_{\rm in}$}
\ee
Typical magnetic field in the non-relativistic sub-layer is 
\be
B_{nr}  \sim  { \delta \over \eta u_{\rm in}} B_{\rm in}  \sim {B_{\rm in} 
\over u_{\rm in}^2} \ll B_{\rm in}
\ee
while the typical current density is 
\be
 j_{nr} \sim { B_{\rm in} \over \eta}  
 \sim {  B_{\rm in} \over \gamma_{\rm in} \delta} 
\ee
Here lies a qualitative difference between the
non-relativistic and relativistic reconnection layer.
In the non-relativistic  Sweet--Parker theory 
the  thickness of  the layer is defined by the 
magnitude of  current density at the center $y=0$.
In  the case of relativistic inflow ($\gamma_{\rm in} \gg 1$),
the current flowing in the bulk  of the
flow $ j \sim  B_{\rm in} / \delta$ is much stronger than 
typical current on the midplane.

The example in this section is an illustration
only, invoked in order to demonstrate 
a possible relation between
$\delta$ and $\beta_{\rm in}$. We use here only Ohm's law to
establish such a connection; in reality,
the velocity profile needs to be determined self-consistently
by solving the entire 2D problem, including the equation of 
motion.

\section{ Relativistic Sweet-Parker model}

In this section we use Ohm's law and the conservation laws
for energy and particle flux and do not solve the  momentum
equation. All the estimates below 
are made up to the order of magnitude, or, more precisely,
how they scale with two parameters, $S$ and~$\sigma$.


Estimating the current in the reconnection layer
\be
j_z \sim {B \over \delta} \sim 
{ \gamma_{\rm in} \beta_{\rm in} B_{\rm in} \over \eta}
\ee
we find
\be
\beta_{\rm in} \gamma_{\rm in} \sim { \eta \over \delta c}
\label{q}
\ee
(we have restored the velocity of light here to make the right-hand 
side explicitly dimensionless).
Introducing  relativistic Lundquist number
\be 
S= { L c \over \eta}  \gg 1
\ee
eq. (\ref{q})
can be rewritten
\be
 \beta_{\rm in}  \gamma_{\rm in}  \sim { L \over \delta} { 1 \over S}
\label{Ohm1}
\ee
This is the first basic equation of the model.

The 
energy flux and the  particle conservation give
\ba &&
\gamma_{\rm in}^2 \beta_{\rm in}  (1+ \sigma ) \rho_{\rm in}^{\ast} L =
 \gamma_{\rm out} ^2  \beta_{\rm out} \rho^\ast_{\rm out}  \delta
\nn  &&
\gamma_{\rm in}  \rho^\ast_{\rm in} \beta_{\rm in} L  =
 \gamma_{\rm out} \rho^\ast_{\rm out}  \beta_{\rm out}  \delta
\ea
In writing down these equations we have assumed that magnetic energy
is fully spent on acceleration of baryons in the downstream flow. 
This assumes that all the electron-positron pairs created in the
reconnection layer have annihilated.

Then it follows that
\be
\gamma_{\rm in} (1+ \sigma) = \gamma_{\rm out}
\label{Gout}
\ee
and
\ba &&
{ \delta \over L } \sim { 1\over  S  \beta_{\rm in} \gamma_{\rm in}}
\nn &&
{ \rho^\ast_{\rm in}  \beta_{\rm in} \over
 \rho^\ast_{\rm out}  \beta_{\rm out} } =
 (1+\sigma) { \delta \over L }
\label{Gs}
\ea

Equation (\ref{Gout}) relates the velocity in the outflowing region to the
inflow velocity and magnetization parameter. 
For small $\sigma \ll 1$  eq. (\ref{Gout})
reproduces the familiar  non-relativistic result that the outflowing
velocity is of the order of the inflowing \Alfven speed:
$\beta_{\rm out} \sim \sqrt{ 2 \sigma } = B/c\sqrt{ 4 \pi \rho}= V_A/c$.
In the relativistic regime,  $\sigma \gg 1$, the outflowing 
velocity  is always relativistic  so that $ \beta_{\rm  out} \sim 1$.
It is remarkable, that in relativistic reconnection the  Lorentz factor 
of the outflowing plasma is {\it much larger} than the  Lorentz factor 
of \Alfven waves in the inflowing region $\gamma_A$:
$\gamma_{\rm out} \sim \sigma \gamma_{\rm in}= \gamma_{A}^2 \gamma_{\rm in}$,
where $ \gamma_{A} \sim \sqrt{\sigma} $. 

Eqns. (\ref{Gout}) and (\ref{Gs}) 
 are general relations for four  unknown quantities:
$\delta, \, \gamma_{\rm in}, \, \gamma_{\rm out}$ 
and $\rho^\ast_{\rm in} / \rho^\ast_{\rm out}$.
In order to resolve this system  we need to use the momentum equations
and find the structure of the flow inside the reconnection region,
 but this is a difficult task. 
An alternative way to proceed is to assume incompressibility of the plasma
  in its rest-frame 
$\rho^\ast_{\rm in} = \rho^\ast_{\rm out}$ (c.f., Blackman \& Field 1994). 
Incompressibility may be due to, for example, a significant
longitudinal
field component  $B_z$.
It is expected that the assumption of incompressibility
will be justified as long as the inflowing plasma velocity is
smaller that the fast magneto-sonic velocity, which in our
case is similar to the  \Alfven  velocity. 

Below we show that for a given value of $S$ the
inflowing velocity increases with the magnetization parameter $\sigma$
reaching at $\sigma  \sim S^2$ the \Alfven  velocity. For such strong
magnetization the compressibility of plasma will become important.
For any $ \sigma  \ll S^2$ the incompressibility is expected to be  a good
approximation.

The assumption of incompressibility 
gives in relativistic limit $\sigma \gg 1$
\be
 \beta_{\rm in} \gamma_{\rm in} L =  \gamma_{\rm out} \delta
\ee
and
\be
\beta_{\rm in} = \sigma { \delta \over L}
\label{beta}
\ee
(here we dropped $\beta_{\rm out}$ because we expect 
an ultra-relativistic outflow with $\beta_{\rm out} \approx 1$.)

Eq. (\ref{beta}) and  the approximate Ohm's law
(\ref{Ohm1}) form a system of two equations for the
thickness of the reconnection layer $\delta$ and the inflow velocity
$\beta _{\rm in}$.
There are two generic  regimes (limiting cases) of relativistic 
reconnection: (i) relativistic inflow $\gamma_{\rm in} \gg 1$ and
(ii) non-relativistic inflow
$\beta_{\rm in} \ll 1$.

\subsection{ Non-relativistic inflow $\beta_{\rm in} \ll 1$ }
Using  eqns. (\ref{beta}) and (\ref{Ohm1})
we find
\be
\beta_{\rm in} \sim { L \over \delta} { 1 \over S} \sim { \sigma \delta \over L}
\ee
Thus,
\be
{\delta \over L} \sim 
{1 \over \sqrt{ S \sigma}} \sim { \beta_{\rm in} \over  \sigma} \sim
{1 \over S \beta_{\rm in}}
\ee
and finally
\be
\beta_{\rm in} \sim \sqrt{ \sigma \over S} \sim \sqrt{ 2 \over S} \gamma_A
\ee
where $\gamma_A = \sqrt{2 \sigma}$ is 
 the Lorentz
factor of the \Alfven wave  velocity in the incoming region.
Since by assumption $\beta_{\rm in} \ll 1$ the 
non-relativistic inflow velocity is realized for
 $\sigma \ll  S$.

Thus, for a given $\sigma$ the inflow velocity is inversely proportional to the
square root of the Lundquist number, similar to the classical non-relativistic 
Sweet-Parker model.

\subsection{Relativistic sub-alfvenic
 inflow $ 1 \ll  \gamma_{\rm in} \ll  \sqrt{2 \sigma}$}

When $\beta _{\rm in} \approx 1$ from eq. (\ref{beta}) it follows
\be 
{ \delta  \over L } \sim {1 \over \sigma}
\ee
and
\ba &&
 \gamma_{\rm out} \sim {  \sigma^2 \over S}
\nn &&
\gamma_{\rm in}  \sim {  \sigma  \over S}
\ea
For consistency we need $ \sigma \gg  S$ (since we have always assumed 
relativistic motion $\gamma \gg 1$).

The ratio of the inflowing  plasma  Lorentz
factor  to the \Alfven wave Lorentz
factor  is 
\be 
{\gamma_{\rm in} \over \gamma_A} \sim {  \sqrt{\sigma /2} \over S}
\ee
Since the inflowing plasma should be sub-alfvenic for the incompressibility
assumption to hold, it is required that
\be
\sigma \ll  2 S^2
\ee
Thus the incompressible  relativistic inflow case is applicable for
\be 
 S \ll \sigma  \ll 2 S^2
\ee

\subsection{ Relativistic alfvenic inflow $\gamma_{\rm in}
\sim \sqrt{2 \sigma} $}

If $  \sigma \geq 2 S^2$ the required inflow velocity becomes
of the order of the \Alfven velocity and the assumption
of incompressibility should break down.
Since the inflow velocity cannot exceed the \Alfven velocity  for
causal reasons, we assume that
 $\gamma_{\rm in} \sim  \gamma_A =  \sqrt{ 2 \sigma}$. 
The parameters of the reconnection layer then become
\ba &&
\gamma_{\rm out} \sim \sqrt{2} \sigma^{3/2}
\nn &&
{\delta \over L } \sim {1\over \sqrt{ 2 \sigma} S}
\nn &&
{ \rho^\ast_{\rm in} 
\over  \rho^\ast_{\rm  out} } \sim 
 \sqrt {  \sigma \over   2}  { 1 \over S} \geq 1
\ea

\section{Bohm diffusion}

In the preceding section we have derived how the flow variables 
 (first of all the inflow velocity) scale with  two parameters of the
model, $S$ and $\sigma$. For astrophysical applications it is often useful
to  have a qualitative estimate of the maximum possible reconnection rate
in a given system. The maximum reconnection rate
corresponds to the maximum value for  resistivity and thus the
minimal Lundquist number. This resistivity  
 may be estimated using Bohm's arguments that the 
maximum diffusion coefficient
in  magnetized plasmas cannot  be much larger than  $r_L v$,
where $ r_L$ is the Larmor radius and $v$ is the typical velocity
of electrons (of the order  of the speed of light in our case).
Thus,
\be
\eta \sim { c^2 \over \om_B}
\ee
and we find
\ba &&
S \sim { L \over r_L}
\label{SS}
\\ &&
\delta \sim {r_L \over \beta_{\rm in} \gamma_{\rm in}}
\label{SS1}
\\ &&
\beta_{\rm in}^2 \gamma_{\rm in} \sim  \sigma {r_L \over L}
\label{SS2}
\ea
Note, that in this case, as $\beta_{\rm in}$ approaches the velocity of light
the  reconnection layer becomes
microscopically thin.
One may expect that the
fluid picture will become inapplicable at this point.

For non-relativistic inflow velocity the assumption of Bohm
diffusion gives
\ba &&
\delta \sim \sqrt{ r_L L \over \sigma}
\nn &&
\beta_{\rm in} \sim  \sqrt{ \sigma r_L  \over  L}
\label{SS4}
\ea
while for relativistic 
 inflow velocity
\ba &&
\delta \sim { r_L  \over  \gamma_{\rm in}}
\nn &&
\gamma_{\rm in} \sim \sigma {r_L \over L}
\label{SS5}
\ea
which requires 
$ \sigma {r_L / L} > 1$.
In the case of  relativistic 
 inflow velocity and under the assumption
of Bohm diffusion  the thickness of the reconnection layer becomes
smaller than the external gyro-radius. 

Equations (\ref{SS2}) and the two limiting cases (\ref{SS4}) and
(\ref{SS5}) give a useful  "order-of-magnitude" estimates of the
potential 
efficiency of reconnection in relativistic plasma.

\section{Discussion}

We have considered the dynamics of the
relativistic Sweet-Parker reconnection under the assumption
that the
inflow region's energy density is dominated by 
magnetic field.
We have found three generic regimes depending on the ratio
of the magnetization parameter $\sigma$ to the Lundquist number $S$:
(i) non-relativistic inflow velocity, $\sigma \ll S$;
(ii) relativistic sub-alfvenic  inflow velocity, $ S  \ll  \sigma \ll 2 S^2$;
(iii)  relativistic alfvenic  inflow velocity, $ \sigma \geq 2 S^2$.
For the first two regimes  plasma flow may be assumed incompressible,
while for the alfvenic  inflow velocity compressibility is important.

 An apparent drawback of our approach is that we did not solve 
in a self-consistent way both 
the momentum  and energy equations. This is a common flaw of many 
models of reconnection based on the Sweet-Parker approach. 
One can say that the role
of the energy equation is to determine how compressible
the plasma is. Conventionally, the  simplest  Sweet--Parker model 
 does not 
include the energy balance equation with all its subtleties 
arising from possible radiation and conducting cooling effects;
instead, it just replaces the energy balance equation with the 
incompressibility condition. In the absence of strong cooling
or when a strong axial magnetic field component is present,
the compressibility effects are indeed not important, at least
in a rough, order-of-magnitude analysis. One then combines
the incompressibility condition with the momentum equation and
Ohm's law to arrive at the Sweet--Parker reconnection scaling.
In our analysis  we also assume incompressibity but then we
use the energy conservation instead of the momentum equation;
in the absence of energy losses these two approaches are, of 
course, equivalent and lead to the same results. 

In spite of the incompressibility assumption our approach 
represents a step forward in understanding relativistic 
reconnection as compared to the view of Blackman \& Field (1994). 
Using the same incompressibility assumption they were able to 
determine only the ratio of inflow and outflow velocities, 
while we find both these quantities separately, expressed 
in terms of external magnetization and the Lundquist number.
We have also found a possible structure of the relativistic 
reconnection layer.

For astrophysical applications, 
we were able to provide order-of-magnitude estimates
of reconnection rates in  relativistic plasma
(eq. \ref{SS2}), which should be used instead of the often {\it ad hoc} 
assumptions of  reconnection rates in  strongly magnetized
plasmas of pulsar winds and AGN jets.
An important result is that under certain conditions the 
inflow velocity  may become relativistic insuring  very
efficient dissipation of magnetic energy.

\begin{acknowledgements}
We would like to thank Eric Priest, Vladimir Pariev, Eric Blackman, 
Chris Thompson
 and  Alissa Nedossekina for comments on the manuscript.
This research was supported  by the NSF grant NSF-PHY99-07949.
DU would like to thank CITA for hospitality during his visit.
\end{acknowledgements}


\appendix

\section{Asymptotic scaling of  $Y(u_{\rm in})$}
\label{Yassym}

In this appendix we investigate the asymptotic behavior of 
the function $Y(u_{\rm in})$ in the case of the 4-velocity
field $u(y)$ described by equation~(\ref{u}), for two limiting 
regimes, $u_{\rm in}\rightarrow 0$ and $u_{\rm in}\rightarrow\infty$.

We start by rewriting equation~(\ref{u}) in terms of rescaled 
coordinate $\tilde{y}=y/\delta$:
\be
{{\partial\hat{B}(\tilde{y})}\over{\partial\tilde{y}}} =
Y \gamma (\beta_{\rm in}-\beta B) \, .
\ee

Now let us integrate this equation from 0 to 1 and use 
the boundary conditions $\hat{B}(0)=0$ and $\hat{B}(1)=1$.
We get
\be
B(1)-B(0) = 1 =
Y \beta_{\rm in} \int\limits_0^1 \gamma(\tilde{y}) d\tilde{y} -
Y \int\limits_0^1 u(\tilde{y}) B(\tilde{y}) d\tilde{y} \, .
\ee

For the velocity field given by equation~(\ref{u}), i.e.
$u(\tilde{y})=u_{\rm in}\tilde{y}$, we then have
\be
Y^{-1} = \beta_{\rm in} \int\limits_0^1 \gamma(\tilde{y}) d\tilde{y}-
u_{\rm in} \int\limits_0^1 \tilde{y} B(\tilde{y}) d\tilde{y} \, .
\label{1/Y}
\ee

Using the relationship $\gamma=\sqrt{1+u^2}$, the integral in 
the first term on the right-hand side can be computed exactly, 
with the result
\be
\int\limits_0^1 \gamma(\tilde{y}) d\tilde{y} = 
{1\over u_{\rm in}} \int\limits_0^{u_{\rm in}} \sqrt{1+u^2} du =
{\sqrt{1+u_{\rm in}^2}\over 2} + 
{1\over 2} \ln (u_{\rm in}+\sqrt{1+u_{\rm in}^2}) \, .
\label{1st-term}
\ee

In the strongly relativistic limit $u_{\rm in}\rightarrow \infty$
this expression can be expanded as 
\be
\int\limits_0^1 \gamma(\tilde{y}) d\tilde{y} =
{u_{\rm in}\over 2} + {1\over 2}\ln u_{\rm in} + O(1) \, .
\label{qQ}
\ee

\begin{figure}[h]
\psfig{file=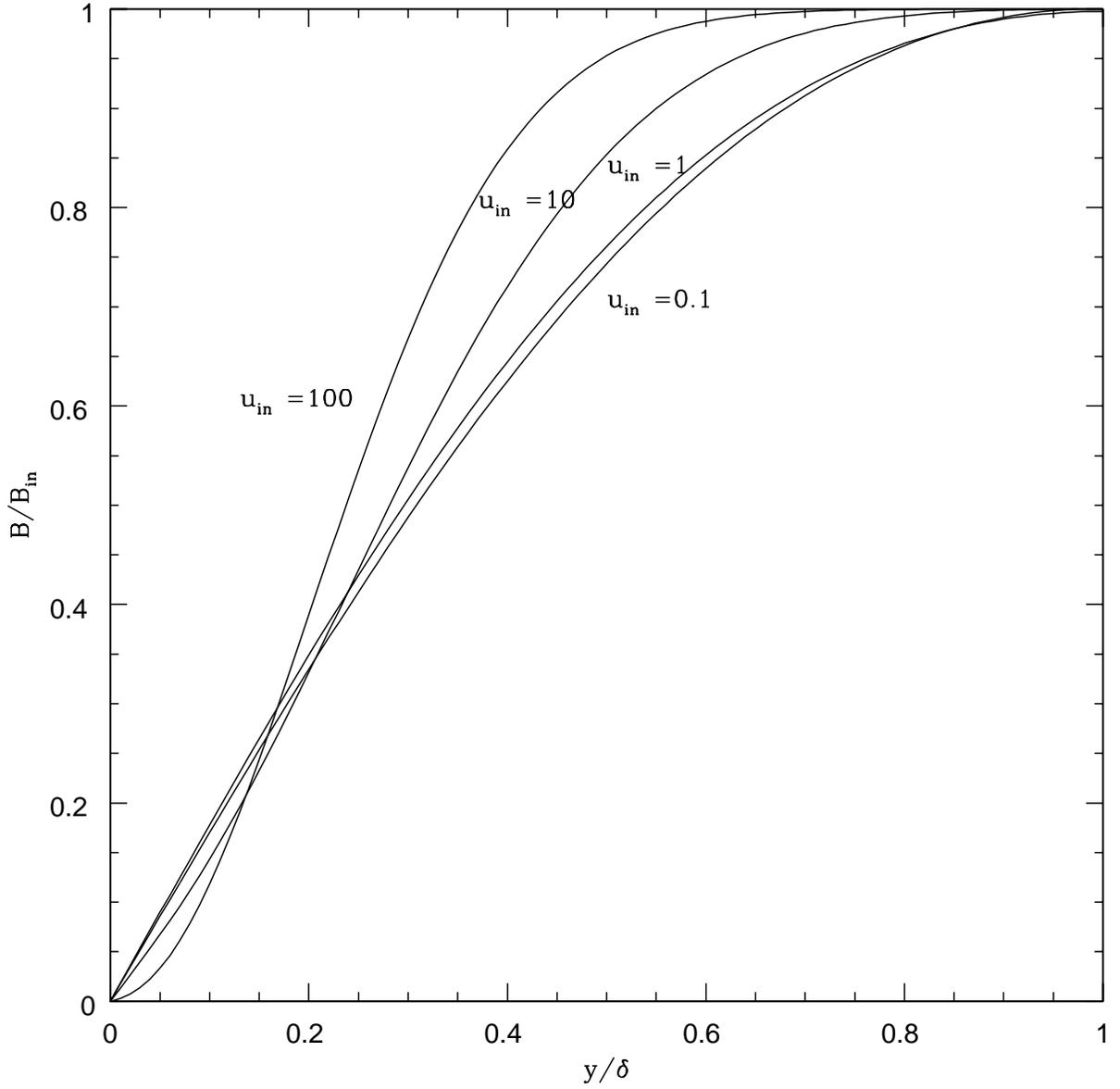,width=17cm}
\caption{Structure of the magnetic field for a linear dependence 
of $u(y)=u_{\rm in} y/\delta$ for different values of $u_{\rm in}$.
}
\label{3}
\end{figure}

As for the second term, we note that in the strongly relativistic 
limit there is a thin non-relativistic boundary sub-layer of 
thickness $\delta_{nr}\sim \delta/u_{\rm in}$; inside this sub-layer 
$\hat{B}(y)$ behaves linearly, with the slope that, as we shall see 
later, is inversely proportional to $u_{\rm in}$. However, what is 
important for us here, outside of this infinitesimally thin sub-layer 
the function $\hat{B}(\tilde{y})$ approaches a certain universal shape, 
$\hat{B}_{\rm rel}(\tilde{y})$ in the limit $u_{\rm in}
\rightarrow \infty$. This conclusion follows from our numerical 
solutions presented in Figure~\ref{3}. Ignoring the non-relativistic
sublayer's contribution to the integral in the second term on the 
right-hand side of equation~(\ref{1/Y}), we can then estimate this 
integral as
\be 
\int\limits_0^1\tilde{y}\hat{B}(\tilde{y})d\tilde{y}\rightarrow 
\int\limits_0^1\tilde{y}\hat{B}_{\rm rel}(\tilde{y}) d\tilde{y}, 
\qquad {\rm as} \quad u_{\rm in} \rightarrow \infty \, ,
\ee

Combining this result with the result (\ref{qQ}) derived for the first term
on the right-hand side of equation~(\ref{1/Y}), we finally can write 
the following expression:
\be 
Y = {1\over{A_{\rm rel} u_{\rm in} + {1\over 2} \ln u_{\rm in} + C}} \, ,
\label{Y_rel}
\ee
where we used the identity $\int_0^1 \tilde{y}d\tilde{y} =1/2$
and defined 
\be
A_{\rm rel} \equiv \int\limits_0^1 
\tilde{y} [1-B_{\rm rel}(\tilde{y})] d\tilde{y} = {\rm const} \, .
\ee

From our numerical solutions with very high (of order $10^3$) 
values of~$u_{\rm in}$, we find $A_{\rm rel}\simeq 0.027$ and $C=-1$.
Thus, we get
\be
Y(u_{\rm in}\rightarrow \infty) \simeq {37\over{u_{\rm in}}}\, 
\left( 1 - {{18.5 \ln u_{\rm in}}\over{u_{\rm in}}} + ... \right) \, .
\ee

\begin{figure}[h]
\psfig{file=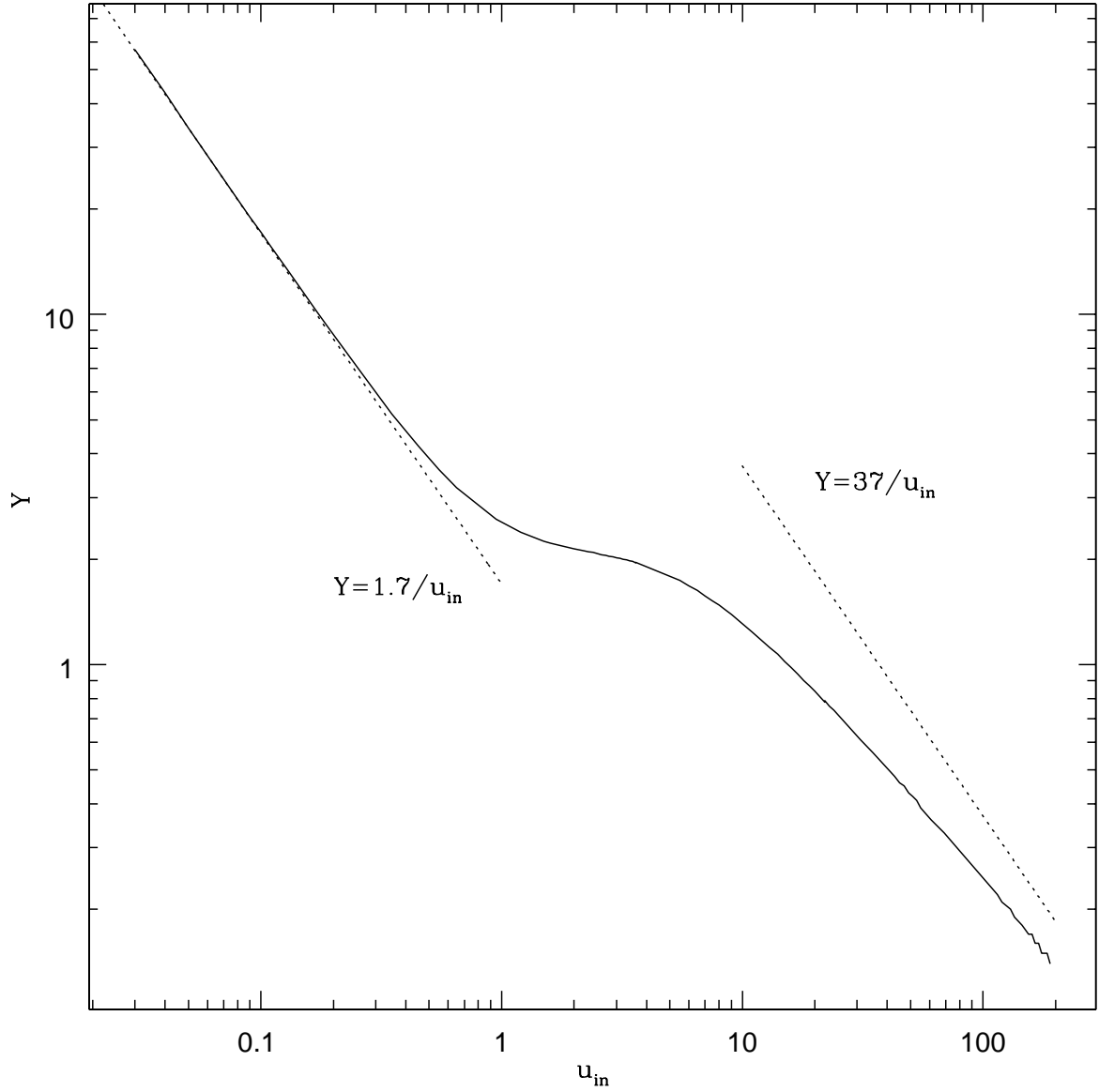,width=17cm}
\caption{Dependence of $Y \equiv \delta/\eta$ on the  four velocity
 of the 
incoming plasma for a linear dependence of $u(y)=u_{\rm in} y/\delta$ 
inside the reconnection layer.
The non-relativistic asymptotic is $ Y = 1.7 / u_{\rm in}$.
}
\label{2}
\end{figure}

\begin{figure}[h]
\psfig{file=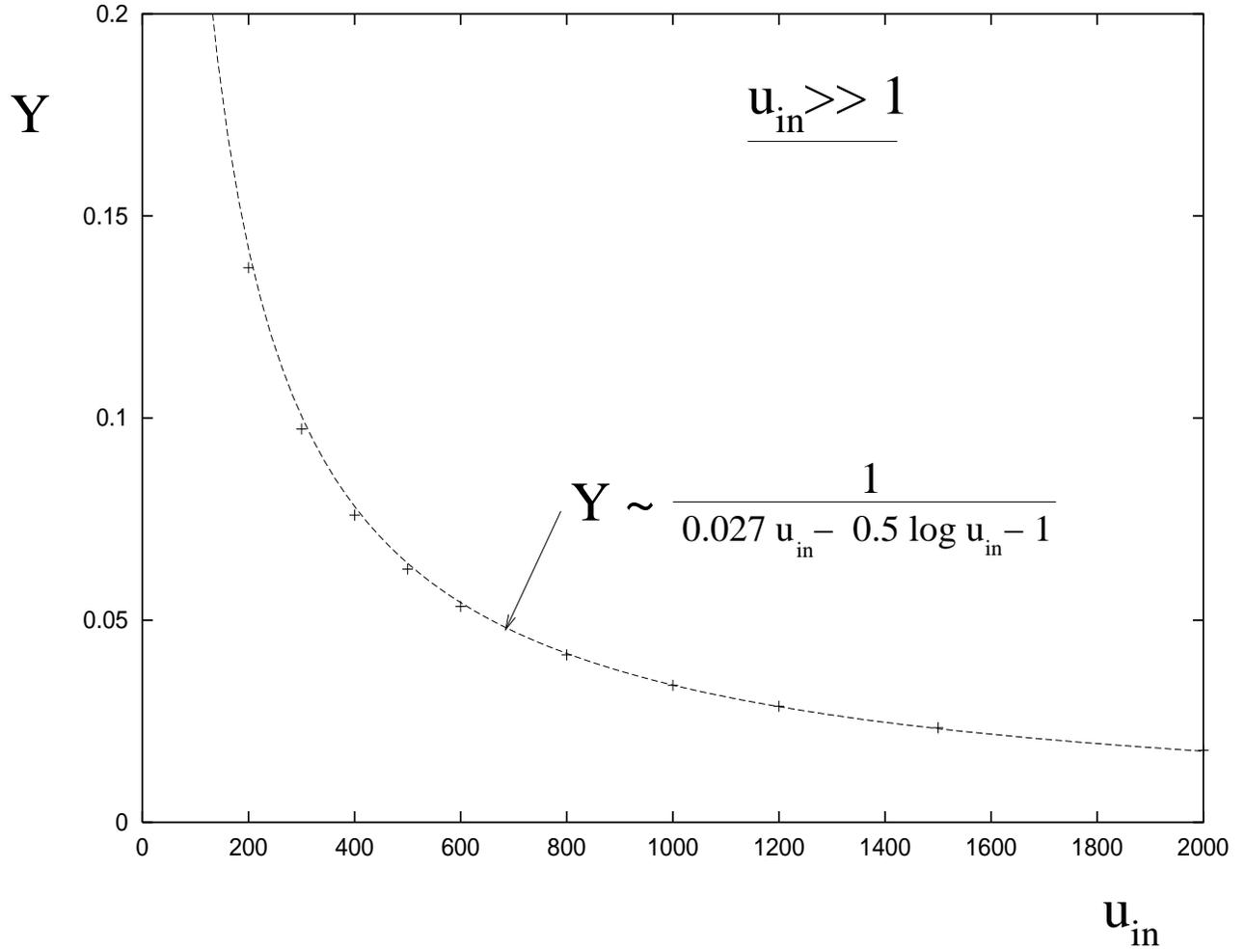,width=17cm}
\caption{Asymptotic behavior of the function $Y(u_{\rm in})$
in the strongly-relativistic limit $u_{\rm in}\rightarrow \infty$.
}
\label{4}
\end{figure}

Notice that, because of the rather large (18.5) numerical coefficient
multiplying the logarithmic factor, this logarithmic correction does
not become negligible until one considers really very large values of
$u_{\rm in}$, of order $10^3$ or more. This is the reason why the plot
on Figure~\ref{2} (with $u_{\rm in}$ up to 200) does not show a very 
good agreement with the simple power law $Y\sim 1/u_{\rm in}$. Notice
however, that the numerical results for values of $u_{\rm in}$ in the 
range between~200 and~2000 are in excellent agreement with the predicted
asymptotic behavior~(\ref{Y_rel}), as demonstrated separately in 
Figure~\ref{4}.

Now let us consider the non-relativistic regime, $u_{\rm in}\ll 1$.
In this limit, the expression~(\ref{1st-term}) for the integral entering the 
first term on the rhs of equation~(\ref{1/Y}) tends to~1 (this is of course 
a trivial result since in this limit $\gamma(\tilde{y})\approx 1$). 
Then, since $\beta_{\rm in}\approx u_{\rm in}$ in this regime, this
first term can be evaluated simply as $u_{\rm in}$.

At the same time, just as in the strongly relativistic case, the function
$\hat{B}(\tilde{y})$ also approaches a certain universal profile,
as can be seen in Figure~\ref{3} (we shall call this profile 
$\hat{B}_{\rm nr}(\tilde{y})$). Then, the integral entering 
the second term on the rhs of equation~(\ref{1/Y}) asymptotically 
approaches a constant value 
\be
\int\limits_0^1 \tilde{y} B(\tilde{y}) d\tilde{y} \rightarrow 
{1\over 2} - A_{\rm nr} = {\rm const} \, ,
\ee
where we defined $A_{\rm nr}$ in a manner similar to $A_{\rm rel}$:
\be
A_{\rm nr} \equiv \int\limits_0^1 
\tilde{y} [1-B_{\rm nr}(\tilde{y})] d\tilde{y} \, .
\ee

Thus, we see that in the non-relativistic limit the function
$Y(u_{\rm in})$ asymptotically approaches an inverse power law:
\be 
Y(u_{\rm in}) \simeq {1\over{u_{\rm in}}}\, {1\over{{1\over 2}+A_{\rm nr}}} 
\simeq {1.71\over{u_{\rm in}}} \, ,
\ee
where we used the value of $A_{\rm nr}=0.0855$ computed using our numerical
solution for the case $u_{\rm in}=0.01$. The resulting asymptotic behavior 
in the non-relativistic regime agrees very well with the numerical results, 
as can be seen in Figure~\ref{2}.

\newpage

\end{document}